\documentstyle[multicol,aps,prl,epsf]{revtex}
\begin{document}
\title{Nontrivial Exponent for Simple Diffusion}
\author{Satya N. Majumdar$^1$, Cl\'ement Sire$^2$, Alan J. Bray$^3$, 
and Stephen J. Cornell$^3$}
\address{
$^1$ Physics Department, Yale University, New Haven, CT 06520-8120 \\
$^2$ Laboratoire de Physique Quantique, Universit\'e Paul Sabatier, 
31062 Toulouse Cedex, France \\
$^3$ Department of Theoretical Physics, The University, Manchester M13 9PL, 
UK. 
}
\date{May 17, 1996}
\maketitle

\begin{abstract}
The diffusion equation $\partial_t\phi = \nabla^2\phi$ is considered, 
with initial condition $\phi({\bf x},0)$ a gaussian random variable with 
zero mean. Using a simple approximate theory we show that the probability 
$p_n(t_1,t_2)$ that $\phi({\bf x},t)$ [for a given space point ${\bf x}$]  
changes sign $n$ times between $t_1$ and $t_2$ has the asymptotic form 
$p_n(t_1,t_2) \sim [\ln(t_2/t_1)]^n(t_1/t_2)^{-\theta}$. The exponent 
$\theta$ has predicted values $0.1203$, $0.1862$, $0.2358$ in dimensions 
$d=1,2,3$, in remarkably good agreement with simulation results.   
\end{abstract}
\begin{multicols}{2}
The diffusion equation, $\partial_t\phi = \nabla^2\phi$, is one of the 
fundamental equations of classical physics. The exact solution of this 
simple equation, for an arbitrary initial condition $\phi({\bf x},0)$, 
can be written down explicitly: 
$\phi({\bf x},t) = \int d^dx\,G({\bf x}-{\bf x}',t)\phi({\bf x}',0)$, 
where $G({\bf x},t) = (4\pi t)^{-d/2} \exp(-x^2/4t)$ is the Green's 
function (or `heat kernel') in $d$ dimensions. The solution is 
characterized by a single growing length scale, the `diffusion length' 
$L(t) \sim t^{1/2}$. It may come as a surprise, therefore, to discover 
that there is a nontrivial exponent associated with this simple process. 

It is the purpose of this Letter to point out that the solutions of the 
diffusion equation exhibit some remarkable and unexpected properties 
associated with their time evolution, and to present a simple theory which 
accounts for this behaviour. We consider specifically a class of initial 
conditions where $\phi({\bf x},0)$ is a gaussian random variable with zero 
mean. Our basic question is the following. What is the probability $p_0(t)$ 
that the field $\phi$ at a particular point ${\bf x}$ has not changed sign 
up to time $t$? Precise numerical simulations in $d=1$ and 2, discussed 
below, demonstrate a power-law decay of the form $p_0(t) \sim t^{-\theta}$, 
with $\theta = 0.1207 \pm 0.0005$ for $d=1$, and $0.1875 \pm 0.0010$ for 
$d=2$. We will present a simple analytic treatment which gives results in 
extraordinarily good agreement with the simulations. Furthermore, 
the analysis gives the more general result 
$p_n(t_1,t_2) \sim [\ln(t_2/t_1)]^n\,(t_1/t_2)^{-\theta}$ for the 
probability that the field changes sign $n$ times between $t_1$ and $t_2$, 
for $t_2 \gg t_1$. The key idea underlying these results is that the gaussian 
process $\phi({\bf x},t)$ is a gaussian {\em stationary} process in terms 
of a new time variable $T=\ln t$. The central assumption in the analysis 
is that the intervals between successive zeros of $\phi({\bf x},T)$ can 
be treated as independent. 

Exponents $\theta$ analogous to that introduced above have recently excited 
much interest in other 
contexts \cite{DBG,BDG,Stauffer,DOS,Boston,DHP,MS,CS,Cardy,BKR}. 
The simplest such system is the $d=1$ Ising model at temperature $T=0$. 
For evolution under Glauber dynamics from a random initial state, the 
probability that a given spin has not flipped up to time $t$ decays as 
$t^{-\theta}$, with $\theta = 3/8$, though the proof of this is surprisingly 
subtle \cite{DHP}. This $d=1$ method is difficult to extend to higher
dimensions, although values for $\theta$ have been obtained 
numerically \cite{DBG,Stauffer,DOS,MS}. An approximate method 
for general dimensions has recently been developed \cite{MS}, whose 
predictions are consistent with simulation results. 
In general, the non-triviality of $p_0(t)$ is a consequence of the
fact that it probes the entire history of a non-Markovian process.
   
We begin by presenting the theoretical approach and the numerical simulation 
results. Experimental ramifications will be discussed briefly at the end of 
the Letter. Other contexts in which a nontrivial exponent $\theta$ might be 
expected will also be discussed. 

The starting point for the discussion of the diffusion equation is 
the expression for the autocorrelation function of the variable 
$X(t)=\phi({\bf x},t)/\langle [\phi({\bf x},t)]^2 \rangle^{1/2}$ for some 
fixed point ${\bf x}$. For `white noise' initial conditions, 
$\langle \phi({\bf x},0)\phi({\bf x}',0) \rangle 
= \delta^d({\bf x}-{\bf x'})$, this takes the form
\begin{equation}
a(t_1,t_2) \equiv \langle X(t_1)X(t_2) \rangle = [4t_1t_2/(t_1+t_2)^2]^{d/4}.
\label{a}
\end{equation}
More generally, this form is asymptotically correct provided the initial 
condition correlator is sufficiently short-ranged (it must decrease 
faster than $|{\bf x}-{\bf x}'|^{-d}$). 

Introducing the new time variable $T=\ln t$, one sees that the 
autocorrelation function becomes $a(T_1,T_2) = f(T_1-T_2)$, where $f(T) = 
[{\rm sech}(T/2)]^{d/2}$. Thus the process $X(T)$ is {\em stationary} 
(the gaussian nature of the process ensures that all higher-order 
correlators are also time-translation invariant). This is an important 
simplification. Note that the anticipated form of the probability of 
$X(t)$ having no zeros between $t_1$ and $t_2$,  
$p_0(t_1,t_2) \sim (t_1/t_2)^\theta$ for $t_2 \gg t_1$, becomes an 
exponential decay, $p_0 \sim \exp[-\theta(T_2-T_1)]$, in the new time 
variable. This reduces the calculation of an exponent to the calculation 
of a decay rate\cite{MS}. The only approximation we shall make is that the
intervals between successive zeros of $X(T)$ are statistically independent.
This `independent interval approximation' (IIA) was introduced in another
context some forty years ago \cite{McFadden}. We shall find that it is an 
extraordinarily good approximation for the diffusion equation. 

As a preliminary step, we introduce the `clipped' variable 
$\sigma = {\rm sign}\,(X)$, which changes sign at the zeros of $X(t)$. 
Clearly, the correlator $A(T) = \langle \sigma(0)\sigma(T) \rangle$ is 
determined solely by the distribution $P(T)$ of the intervals between zeros.  
The strategy is to determine $P(T)$ from $A(T)$, and $p_0$(T) from $P(T)$. 
To this end we note first that 
\begin{equation}
A(T)=\frac{2}{\pi}\sin^{-1}[(a(T)] = 
\frac{2}{\pi}\sin^{-1}\left([\rm{sech}\,(T/2)]^{d/2}\right),
\label{A}
\end{equation}
where the first equality holds for any gaussian process.  

Next one expresses $A(T)$ in terms of the interval-size distribution $P(T)$. 
Clearly
\begin{equation}
A(T) = \sum_{n=0}^{\infty} (-1)^n p_n(T),
\label{Auto}
\end{equation}
where $p_n(T)$ is the probability that the interval $T$ contains $n$ zeros 
of $X(T)$. We define $Q(T)$ to be the probability that an interval of size $T$ 
to the right or left of a zero contains no further zeros. Then $P(T)=-Q'(T)$. 
For $n \ge 1$ one obtains immediately
\begin{eqnarray}
p_n(T) = &&\langle T \rangle^{-1}\int_0^T dT_1 \int_{T_1}^T dT_2 \ldots 
\int_{T_{n-1}}^T dT_n\times \label{intervals}\\
&& \times Q(T_1)P(T_2-T_1)\ldots P(T_n-T_{n-1})Q(T-T_n),
\nonumber
\end{eqnarray}
where $\langle T \rangle$ is the mean interval size. One has made the IIA 
by writing the joint distribution of $n$ successive zero-crossing intervals
as the product of the distribution of single intervals. 
Taking Laplace transforms gives $\tilde{p}_n(s)=[\tilde{Q}(s)]^2
[\tilde{P}(s)]^{n-1}$/$\langle T \rangle$. But $P(T)=-Q'(T)$ implies 
$\tilde{P}(s)=1-s\tilde{Q}(s)$, where we have used $Q(0)=1$. Using this 
to eliminate $\tilde{Q}(s)$ gives the final result
\begin{eqnarray}
\tilde{p}_n(s) & = & \frac{1}{\langle T \rangle s^2}
\left(1-\tilde{P}(s)\right)^2 \left(\tilde{P}(s)\right)^{n-1},
\ \ \ n \ge 1, \label{prob_n} \\
& = & \frac{1}{\langle T \rangle s^2}
\left(\langle T \rangle s - 1 + \tilde{P}(s)\right),\ \ \ n=0,
\label{prob_0}
\end{eqnarray}
where the result for $\tilde{p}_0(s)$ follows from the normalization condition 
$\sum_{n=0}^\infty p_n(t) = 1$, which gives $\sum_{n=1}^\infty \tilde{p}_n(s)
=1/s$.

Finally the Laplace transform of (\ref{Auto}) gives $\tilde{A}(s)=
\sum_{n=0}^\infty (-1)^n \tilde{p}_n(s)$. Performing the sum employing  
(\ref{prob_n}) and (\ref{prob_0}), and using the result to express 
$\tilde{P}(s)$ in terms of $\tilde{A}(s)$ gives the desired result
\begin{equation}
\tilde{P}(s) = [2-F(s)]/F(s),
\label{P}
\end{equation}
where
\begin{equation}
F(s) = 1 + (\langle T \rangle/2)\,s[1 - s\tilde{A}(s)].
\label{F}
\end{equation}

Equations (\ref{prob_n}-\ref{F}) are a general consequence of the 
independent interval approximation. The function $F(s)$, defined by 
(\ref{F}), is completely determined by the autocorrelation function 
$A(T)$, and contains all the information needed to compute the 
probabilities $p_n(T)$. We have in mind, of course, to apply this 
approach to the diffusion equation, where $A(T)$ is given by (\ref{A}). 
For this case the mean interval size $\langle T \rangle$, required in 
(\ref{F}), can be simply evaluated. For $T \to 0$, the probability to 
find a zero in the interval $T$ is just $T/\langle T \rangle$, so 
$A(T) \to 1 - 2T/\langle T \rangle$. This gives $\langle T \rangle 
= -2/A'(0) = \pi\sqrt{8/d}$, using (\ref{A}) in the final step.

We note a very important point at this stage. The fact that $A'(0)$ is
finite (i.e.\ $f'(0)=0$ and $f''(0)\neq 0$) is special to the diffusion
equation, which allows us to use the IIA. Physically this means that the 
density of zeros is a finite number. However, for many Gaussian stationary 
processes, such as the one that arises in an approximate treatment of 
the Ising model \cite{MS}, $f'(0)\neq 0$, implying that $A'(0)$ diverges. 
In this case, the IIA cannot be used. For such processes, the variational 
and perturbative methods developed in Ref.\cite{MS} give reasonably 
accurate results.

The asymptotics of $p_0(T)$ are controlled by the singularity of 
$\tilde{p}_0(s)$ with the largest real part, i.e.\ [from (\ref{prob_0})] 
by the corresponding singularity of $\tilde{P}(s)$. The expectation that 
$p_0(T) \sim \exp(-\theta T)$ suggests that this singularity is a simple 
pole, i.e.\ that $F(s)$ has a simple zero at $s=-\theta$. Using (\ref{A}) 
in (\ref{F}), and inserting $\langle T \rangle = \pi\sqrt{8/d}$, gives
\begin{eqnarray}
F(s) = 1 + \pi\left(\frac{2}{d}\right)^{1/2} 
s\left[1 - \frac{2s}{\pi} \int_0^\infty dT\,\exp(-sT)\right.\times&&\nonumber\\
\times\left.\sin^{-1}
\left({\rm sech}^{d/2}\left(\frac{T}{2}\right)\right)\right]&&
\label{F(s)}
\end{eqnarray}
Clearly $F(0)=1$, while $F(s)$ diverges to $-\infty$ for $s \to -d/4$. 
Between these two points $F(s)$ is monotonic, implying a single zero in 
the interval $(-d/4,0)$. Solving (\ref{F(s)}) numerically for this zero, 
and identifying the result with $-\theta$, gives the values of $\theta$ 
shown in table 1. For future reference, we note from (\ref{P}) that the 
residue $R$ of the corresponding pole of $\tilde{P}(s)$ is $R=2/F'(-\theta)$. 
The values of $R$, which controls the amplitude of the asymptotic decay 
of $p_n(T)$, are also given in table 1. Recall that the behaviour 
$p_0(T) \sim \exp(-\theta T)$ translates in `real' time to a decay law 
$p_0(t) \sim t^{-\theta}$ for the probability that $\phi$ at a given 
point has not changed sign. It is also easy to extract the large-$d$
behaviour of $\theta$ from Eq. (9): we find, to leading order in $d$,\ 
$\theta \approx 0.145486 {\sqrt d}$. 

The predicted values of $\theta$ were tested in $d=1$ and 2 by numerical 
simulations.  The diffusion equation was discretized in space and time 
in the form 
\begin{equation}
\phi_i(t+1) = \phi_i(t) + a \sum_j[\phi_j(t) - \phi_i(t)],
\end{equation}
where $j$ runs over the nearest neighbours of $i$ on a linear ($d=1$) or square 
($d=2$) lattice. A stability analysis shows that the solution is unstable for 
$a \ge a_c = 1/(2d)$. Preliminary studies showed that the asymptotic exponent 
is independent of $a$ for $a<a_c$, but that a value $a=a_c/2$ seems to give 
the quickest onset of the asymptotic behaviour. This value was therefore used 
in all simulations reported here. Systems of $2^{20}$ ($2^{22}$) sites were 
studied in $d=1$ ($d=2$), for times up to $2^{17}$ ($2^{12}$). 
Data for longer times in $d=2$ suffer from noticeable finite-size 
effects. The initial values of $\phi_i$ were chosen independently from a 
gaussian distribution of zero mean. Using a rectangular distribution 
gave the same asymptotic exponent within the errors. Several random 
number generators were tried: All gave consistent results (within the 
errors). 
 
The simulation results are presented in Figure 1. The data are an average 
of 17 ($d=1$) and 22 ($d=2$) runs with independent initial conditions. 
An effective exponent $\theta(t)$ is extracted from a least-squares fit of 
$\log_2 p_0$ against $\log_2 t$ over 5 consecutive values of $\log_2 t$. 
The error bars shown in the figure were obtained from the fits. The resulting 
exponent $\theta(t)$ is then plotted against $1/\log_2 t$, where here 
$\log_2 t$ is the mid-point of the 5 values. The best estimates of $\theta$, 
shown in table 1, were obtained by plotting $t^\theta p_0(t)$ against 
$\log_2 t$ and choosing $\theta$ such that, after an initial transient, the 
data show no systematic upward or downward trend with increasing $t$. The 
agreement with the theoretical predictions (table 1) is quite remarkable, 
showing that the IIA is an extraordinarily good approximation in this 
context.
 
The case of correlated initial conditions is also of interest. If the 
Fourier-space correlations are $\langle \phi_{\bf k}(0)\phi_{-\bf k}(0)
\rangle \sim k^\sigma$ for $k \to 0$ ($\sigma > -d$), the autocorrelation 
function of $X(t)$ still has the form (\ref{a}), but with $d$ replaced 
by $d+\sigma$. The independent interval approximation, therefore, predicts 
that the dependence of $\theta$ on $d$ and $\sigma$ enters only through 
the combination $d+\sigma$. To test this, we simulated the case $d=1$, 
$\sigma=2$, noting that $\sigma=2$ corresponds in real space to 
differentiating uncorrelated initial conditions (or taking finite 
differences on a lattice). The result from 12 runs (Figure 1) is 
$\theta = 0.2380 \pm 0.0015$, close (as anticipated) to the predicted 
result $0.2358$ for uncorrelated initial conditions in $d=3$.

The asymptotics of the probability $p_n(t_1,t_2)$ for having $n$ zeros  
between times $t_1$ and $t_2$ are also readily calculable within the 
IIA. From (\ref{prob_n}) and (\ref{prob_0}), the singularity in 
$\tilde{p}_n(s)$ as $s=-\theta$ is an $(n+1)^{\rm th}$-order pole of 
strength $R^{n+1}/\langle T \rangle \theta^2$, where $R$ is the strength 
of the simple pole in $\tilde{P}(s)$. Inverting the Laplace transform, 
and retaining only the leading large-$T$ behaviour, gives (for all $n$)
\begin{equation}
p_n(T) \to \frac{R}{\langle T \rangle \theta^2}\,\frac{(RT)^n}{n!}\,
\exp(-\theta T).
\end{equation}
With $T=\ln(t_2/t_1)$, one obtains 
\begin{equation}
p_n(t_1,t_2) \to (R^{n+1}/\langle T \rangle \theta^2)\,
[\ln(t_2/t_1)]^n\,(t_1/t_2)^\theta.
\end{equation}
When the time $t_1$ corresponds to the initial condition, one has to set 
$t_1$ equal to a constant of order unity, as was implicit in the earlier 
treatment of $p_0(t)$. Setting $t_2=t$ one then gets 
$p_n(t) \sim (\ln t)^n\,t^{-\theta}$. This rather strange-looking result 
does not have the scaling form found in the voter model and in Ising 
systems in $d=1$ and 2, where one finds \cite{Boston,CB} 
$p_n(t) \sim \langle n \rangle^{-1} f(n/\langle n \rangle)$, with 
$\langle n \rangle \sim \sqrt{t}$. [The exponent $\theta$ in those 
systems emerges from a singular behaviour of the scaling function 
$f(x)$ as $x \to 0$. Note that in the present work 
$\langle n \rangle \sim T \sim \ln t$.] 

We turn to a brief discussion of the experimental relevance of our 
results. The ubiquity of the diffusion equation in physics implies that 
applications will be many and varied. As a concrete example, however, 
consider the reaction-diffusion process $A + B \to C$, where $C$ is inert 
and immobile. The corresponding rate equations for the concentrations 
are $dn_A/dt = \nabla^2 n_A - R$, $dn_B/dt = \nabla^2 n_B - R$, 
and $dn_C/dt = R$, where $R$ is the reaction rate per unit volume 
($R \propto n_An_B$ for $d>2$ \cite{RD1}). The concentration 
difference, $\Delta n \equiv n_A - n_B$, obeys the simple diffusion 
equation. If the $A$ and $B$ species are randomly mixed at $t=0$ 
the system evolves, for $d<d_c=4$, to a coarsening state in which the two 
species segregate into domains \cite{R-D}, separated by domain walls whose 
locations are defined by $\Delta n=0$. Subsequent production of the inert 
species $C$ is slaved to the motion of the domain walls, which are zeros of 
the diffusion field $\Delta n$. The fraction of space not infected by the 
$C$ species will therefore decay asymptotically as $t^{-\theta}$. 
\begin{figure}[b]
\narrowtext
\epsfxsize=\hsize
\epsfbox{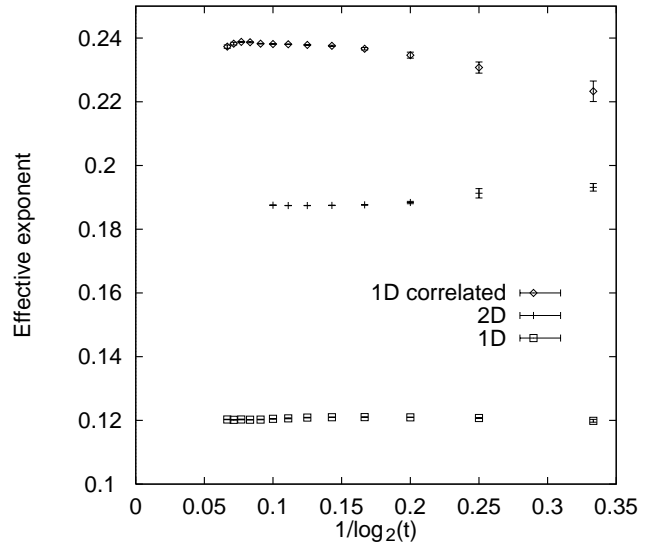}
\caption{Effective exponents $\theta(t)$ plotted against 
$1/\log_2t$ for the diffusion equation in $d=1$ (lower data set), $d=2$ 
(middle set), and $d=1$ with correlated initial conditions (upper set). 
The downturn in the upper set at late times is not statistically 
significant (note the larger errors on the last two points). The best 
estimates of $\theta$ are given in table 1.}
\end{figure}
We conclude with other examples of non-trivial exponents $\theta$ 
which have not been addressed in the literature. The first is associated 
with the dynamics of the {\em global} order parameter $M(t)$ (e.g.\ the 
total magnetization of an Ising ferromagnet) at a critical point $T_c$, 
following a quench to $T_c$ from the high-temperature phase. The quench 
prepares the system in a state with random initial conditions. In the 
subsequent evolution (now stochastic, rather than deterministic), the 
probability that $M(t)$ has not changed sign since $t=0$ decays as 
$t^{-\theta_c}$, where $\theta_c$ is a {\em new critical exponent} 
\cite{MBC}. For reasons similar to those given for the diffusion 
problem, we expect $\theta_c$ to be an independent exponent, i.e.\ 
not related by any scaling law to the usual static and dynamic exponents. 
As a second example, one can consider $M(t)$ for a quench to $T=0$ from 
high temperature. In this case, $p_0(t) \sim t^{-\theta_0}$, where 
$\theta_0$ differs from the corresponding exponent for single spins. 
For the $d=1$ Glauber model, for example, the probability that $M(t)$ 
has not changed sign decays with an exponent $\theta_0 = 1/4$ \cite{MBC}, 
which differs from the exponent 3/8 obtained for the zero-flip 
probability of a given spin \cite{DHP}. 

As a final example, consider the generalised one-dimensional random-walk 
equation $d^n x/dt^n = \xi(t)$, where $\xi$ is gaussian white noise. The 
cases $n=1,2,\dots$ correspond to a random velocity (the usual random walk), 
random acceleration, etc. The first two $\theta_n$ are $\theta_1=1/2$ and 
$\theta_2=1/4$ \cite{theta2}, but larger $n$ have not been considered 
before to our knowledge. Application of the independent interval 
approximation \cite{BCMS} gives equations of the same structure as for the 
diffusion process, but with ${\rm sech}^d(T/2)$ in (\ref{A}) and (\ref{F(s)}) 
replaced by $(2-1/n)\exp(-T/2)\,_2F_1[1,1-n;1+n;\exp(-T)]$, where $_2F_1$ 
is the hypergeometric function. This approach gives $\theta_2 = 0.2647$ 
(instead of $1/4$) while, for larger $n$, $\theta_n$ approaches a limiting 
value $\theta_\infty = 0.1862\ldots$, i.e.\ the same exponent as the $d=2$ 
diffusion equation! In fact, the equality of the exponents for the 
$n=\infty$ process and $d=2$ diffusion can be proved exactly \cite{BCMS}, 
implying a limiting exponent $0.1875 \pm 0.0010$ (from table 1) for the 
former.

To summarize, we have calculated the probability for $n$ zero crossings,  
between times $t_1$ and $t_2$, of a diffusion field at a given point in 
space, by assuming that the intervals between crossings, measured 
in the variable $T=\ln t$, are independent. The time-dependence of these 
probabilities is characterized by a single non-trivial exponent $\theta$, 
the predicted values of which are in excellent agreement with precise 
simulation results in one and two dimensions. These ideas are relevant
to any system where the diffusion equation (or `heat equation') plays a 
role, ranging from physical and chemical systems to fluctuations in 
financial markets, and can be extended to other gaussian processes. 

We thank the Parallel Computing Center of the University of Geneva for 
time on the Connection Machine CM200. AB and SC's research was supported 
by EPSRC (UK), S.M.'s by NSF grant no. DMR-92-24290. We thank I. Gruzberg, 
T. Newman and S. Sachdev for useful discussions. After the manuscript was 
completed, we learned of related work by V. Hakim, B. Derrida, and 
R. Zeitak \cite{HDZ}.  
\begin{table}
\narrowtext
\begin{tabular}{cccc}
$d$&$\theta_{th}$ 
& $\theta_{sim}$
& $R$ \\
\hline 
1 & 0.1203 & 0.1207 $\pm$ 0.0005 & 0.1277 \\
2 & 0.1862 & 0.1875 $\pm$ 0.0010 & 0.2226 \\
3 & 0.2358 & \phantom{$^*$}0.2380 $\pm$ 0.0015$^*$ & 0.2940 \\
4 & 0.2769 &  -- & 0.3527 \\
5 & 0.3128 &  -- & 0.4033 \\
\end{tabular}
\caption{Exponents $\theta$ from theory ($\theta_{th}$) and 
simulations ($\theta_{sim}$), and the value of the residue R (see text), 
for various spatial dimensions $d$. The `$d=3$' simulation result 
($^*$) refers to a $d=1$ simulation with correlated initial conditions 
(see text).}
\end{table}

\end{multicols}


\begin{references}

\bibitem{DBG} B. Derrida, A. J. Bray, and C. Godr\`eche, 
J. Phys.\ A {\bf 27}, L357 (1994). 
\bibitem{BDG} A. J. Bray, B. Derrida, and C. Godr\`eche, Europhys.\ Lett.\ 
{\bf 27}, 175 (1994).
\bibitem{Stauffer} D. Stauffer, J. Phys.\ A {\bf 27}, 5029 (1994). 
\bibitem{DOS} B. Derrida, P. M. C. de Oliveira, and D. Stauffer, 
Physics A {\bf 224}, 604 (1996). 
\bibitem{Boston} E. Ben-Naim, L. Frachebourg, and P. L. Krapivsky, 
Phys. \ Rev.\ E {\bf 53}, 3078 (1996).  
\bibitem{DHP} B. Derrida, V. Hakim, and V. Pasquier, Phys.\ Rev.\ Lett.\ 
{\bf 75}, 751 (1995). 
\bibitem{MS} S. N. Majumdar and C. Sire, preprint {\em cond-mat/9604151} 
(1996). 
\bibitem{CS} S. J. Cornell and A. J. Bray, preprint {\em cond-mat/9603143} 
(1996).
\bibitem{Cardy} J. Cardy, J. Phys. \ A, {\bf 28}, L19 (1995). 
\bibitem{BKR} E. Ben-Naim, P. L. Krapivsky, and S. Redner, Phys.\ Rev.\ E 
{\bf 50}, 2474 (1994). 
\bibitem{McFadden} J. A. McFadden, {IRE Transaction on Information Theory}, 
IT-4, p.14 (1957). 
\bibitem{CB} S. J. Cornell and A. J. Bray, unpublished. 
\bibitem{RD1} S. J. Cornell and M. Droz, Phys.\ Rev.\ Lett.\ {\bf 70}, 
3824 (1993); B. P. Lee and J. Cardy, J.\ Stat.\ Phys.\ {\bf 80}, 971 (1995).
\bibitem{R-D} D. Toussaint and F. Wilczek, J. Chem.\ Phys.\ {\bf 78}, 
2642 (1978). 
\bibitem{MBC} S. N. Majumdar, A. J. Bray, S. J. Cornell, and C. Sire, 
unpublished. 
\bibitem{theta2} T. W. Burkhardt, J. Phys.\ A {\bf 26}, L1157 (1993). 
\bibitem{BCMS} A. J. Bray, S. J. Cornell, S. N. Majumdar, and C. Sire, 
unpublished. 
\bibitem{HDZ} V. Hakim, B. Derrida, and R. Zeitak, unpublished. 
\end{references}
\end{document}